\input harvmac
\input epsf
\Title{\vbox{\hbox{
HUTP-98/A038}\hbox{hep-th/9803249}}}
{\vbox{\hbox{\centerline{Large N limit of {\it orbifold} field theories}}}}
\centerline{Michael Bershadsky and Andrei Johansen}
\centerline{\it Lyman Laboratory of Physics, Harvard
University, Cambridge, MA 02138}
\vskip .5in

\centerline{\bf Abstract}
We consider certain orbifoldization of the ${\cal N}=4$
field theories that leads to  ${\cal N}=2,1,0$ field theories in 4 dimensions.
These theories were recently analyzed using the string 
theory perturbation 
technique.
It was found that in the large $N$ limit all correlation 
functions of the 
orbifold theories coincide with those of ${\cal N}=4$, modulo the rescaling of 
the 
gauge coupling constant. In this paper we repeat the same analysis using the 
field theoretical language.

\vskip .2in
\Date{March 27, 1998}


Recently, it was a considerable interest in constructing conformal 
field theories with small number of supersymmetries ${\cal N}=2,1,0$ 
\ref\kach{S.  Kachru   and  E.  Silverstein,
{\it  4d Conformal Field Theories and Strings on Orbifolds}, hep-th/9802183.}
 \ref\lnv{A. Lawrence, N. Nekrasov and C. Vafa,
{\it On Conformal Theories in Four Dimensions},
 hep-th/9803015.}.
The construction was inspired by an orbifold procedure very well familiar 
from the string theory. We are going to call these theories ``orbifold'' field 
theories,
keeping in mind that orbifoldization is really well understood in the context 
of string theory. As it turned out later the orbifold field theories
were conformal only in the large N limit. Moreover, all correlation functions of 
the 
orbifold theories coincide with those of ${\cal N}=4$, modulo the rescaling of 
the 
gauge coupling constant. For the first glance this statement sounds completely 
shocking.
Namely, all the complicated (and certainly unknown) momenta dependence
of ${\cal N}=2,1,0$ correlation functions should be exactly the same as in 
${\cal N}=4$
theory. 
This was shown using the string theory methods 
\ref\bkv{M. Bershadsky, 
Z. Kakushadze and C. Vafa,
{\it String Expansion as Large N Expansion of Gauge Theories},
 hep-th/9803076\semi
Z. Kakushadze,
{\it Gauge Theories from Orientifolds and Large N Limit}, 
hep-th/9803214.}. 
In fact the string theory methods appear to be very useful in summing the 
Feynman diagrams in various field theories \ref\kos{Z. Bern and D.A. Kosower, Phys. Rev. Lett. {\bf 66} (1991) 1669}(for a
recent discussion, see, {\it e.g.}, \ref\mreof{
Z. Bern, L. Dixon, D.C. Dunbar, M.Perelstin and J.S. Rozowsky, ``On the Relationship between Yang-Mills Theory and Gravity and its Implication for Ultraviolet Divergences'', hep-th/9802162.}).
Still the field theory proof of this statement seems to
be desirable. Moreover, the string theory proof is literally valid only
for the scattering of on-shell particles. As we will see in case of the field 
theory one can also prove the off-shell statement.

We start our discussion by reviewing orbifoldization procedure.
The orbifold group $\Gamma$ is a discrete subgroup of the 
${\cal N}=4$ R-symmetry group $SU(4)$. Depending on whether $\Gamma$
lies entirely in $SU(2)$, $SU(3)$ or in $SU(4)$ 
one gets ${\cal N}=2,1,0$ supersymmetric field theories. 
One also has to specify a representation of $\Gamma$
in $SU(|\Gamma| N)$ group
\eqn\rep{g \in \Gamma ~:~~~ g \rightarrow \gamma_g~,}
where $\gamma_g$ acts by conjugation.
The choice of $\gamma$-matrices determines the embedding the $\Gamma$ in the 
gauge symmetry group.
The consistency of string theory implies that the representation of 
$\Gamma$ has to be {\it regular} -- ${\rm Tr}(\gamma_g)=0$ for all $g \neq {\rm 
I}$.
The importance of regular representations is not so clear from the point of 
view  of field theory. If one relaxes the condition of regularity
then in certain cases the orbifold theory appears to be anomalous, 
but not necessarily. Below we will see how the analysis of field theory perturbation 
technique simplifies for regular representations.

The truncation ($=$orbifoldization) procedure has a bigger realm of applications.  One can start with any gauge theory, not necessarily 
supersymmetric (or conformal) and truncate it according to the action of some
discrete group $\Gamma$. Now $\Gamma$ is an arbitrary discrete group
embedded into the gauge group (via \rep) and non-anomalous flavour group. 
If the representation of $\Gamma$ is regular, then in the large N limit
all the diagrams (of invariant fields) are identical in both theories modulo some rescaling.
The proof presented in this paper is literally applicable to the case when the matter in the original theory is in the adjoint representation and 
can be easily  modified in case of other representations.

The spectrum of the orbifold theory is 
determined by the truncation  
procedure.
The states that survive  in the orbifold 
theory are are those, invariant under the action of $\Gamma$. 
By contrast with the string theory, there are no new states
(orbifold sectors). 
All states come from the original ``parent'' theory.
The projector $P$ to the invariant states 
can be defined as follows
\eqn\proj{P={1 \over  |\Gamma|} \sum_{g \in \Gamma} r_g\otimes 
\gamma_g^+ \otimes \gamma_g ~,}
where $r_g$ denotes a certain representation of 
an element $g\in \Gamma$ in the $R$-symmetry group.
Indeed, it is easy to check that
\eqn\invar{\eqalign{
(r_b\otimes \gamma_b^+\otimes \gamma_b)~P=
{1\over |\Gamma|}\sum_{g\in \Gamma} 
(r_b\otimes \gamma_b^+\otimes \gamma_b)\cdot 
(r_g\otimes \gamma^+_g\otimes \gamma_g)=\cr
{1\over |\Gamma|}\sum_{g\in \Gamma}  
(r_br_g\otimes \gamma_b\gamma^+_g\otimes \gamma_g\gamma_b)=
{1\over |\Gamma|}\sum_{c\in \Gamma} 
(r_c\otimes \gamma_c^+\otimes \gamma_c)=P~.}}
Here we used the fact that multiplication by an element 
$g$ of
the group $\Gamma$
acts to the group as permutation.
Therefore the summing over $a$ can be turned into the summing over 
$c=bg$.
This property ensures that $P$ is indeed projector, namely $P^2=P$.

The choice of representation of $\Gamma$, and hence the explicit
form of the projector $P$ depends on spins 
of the  fields.
In particular
the vector particles in the orbifold theory are singlets and they 
satisfy the condition
\eqn\vec{A = \gamma_g^+ A \gamma_g  ~~~{\rm for}~{\rm any}~g~.}
The matter spectrum can also be determined from a 
similar considerations. 
The adjoint ${\cal N}=4$ scalars transforms in 
representation ${\bf 6}$ with respect to 
R-symmetry and therefore the projection for the 
matter appears to be different
\eqn\bos{\Phi_i = \gamma_g^+ ( (r_g ^{\bf 6})_i ^j \Phi_j ) \gamma_g  ~~~{\rm 
for}~{\rm any}~g~.}
There is also a simple modification of this projection for Weil fermions.
The spectrum of the orbifold theory can be encoded in the quiver diagram. 
This diagram consists of nodes (one per irreducible representation of $\Gamma$)
and some number of fermionic and bosonic arrows connecting them.
With each arrow one associate a matter in bifundamental representation
(see \lnv\ for details).

Now let start the analysis of the Feynman diagrams. 
The external lines are 
restricted
to the orbifold theory. 
In the large N limit  the propagator of the adjoint fields can be written as
\eqn\prop{\langle X_{ij}(p)  X_{kl} (-p)    \rangle  = \delta_{ik} \delta_{jl} 
f(p)}
It is natural to draw this propagator using a double line (or a strip)
(for details see 't Hooft \ref\thoof{G 't Hooft,
{\it A Planar Diagram Theory For Strong Interactions},
Nucl. Phys. {\bf 72} (1974) 461.}).
Here $X_{ij}$ stands for any kind of field in the theory --
vectors, scalars of fermions.

In order to construct the ${\cal N}=2,1,0$
diagram out of ${\cal N}=4$ diagram one has to insert a projector \proj\ in 
every propagator.
It is important that interaction vertices are invariant under 
$\Gamma$ because the vertices of the ${\cal N}=4$ theory 
are invariant under the $SU(4)$
R-symmetry. 
Therefore they do not require any additional projection.

Below, we compare the Feynman diagrams of ${\cal N}=4$ theory with the 
Feynman diagrams of the orbifold theory. ${\cal N}=4$ R-symmetry does not
commute with ${\cal N}=1$ superformalism and therefore one has to consider 
Feynman diagrams for all fields separately.  We will also choose the  gauge
fixing conditions for orbifold theory to be consistent with the 
gauge fixing conditions of ${\cal N}=4$ theory.
As the result the diagrams involving  ghosts of the orbifold 
theory appears by the reduction from ${\cal N}=4$.

Consider a planar Feynman diagram with $K$ external lines and  $R$  propagators.
After replacing every propagator by a strip 
and contracting the indices (meaning 
glueing
the strips together in a ribbon diagram) one 
gets a two dimensional surface with 
$L+1$ boundaries (there are $L$ internal boundaries without
external lines and one boundary with attached external lines). 
A planar diagram has a topology of a disk with $L+1$ boundaries. 
The number of internal boundaries coincide with the number of loops in the  diagram.
In order to get a diagram  in the orbifold theory one 
would insert $R$ 
projectors -- one per propagator. As the result the diagram 
can be written as a sum
\eqn\facr{{1 \over |\Gamma|^R} \sum_{g_1,g_2,...g_r \in \Gamma} 
F_{g_1...g_r}(p_1,...p_K)}
Functions $F_{g_1...g_r}(p_1,...p_K)$ depends on the choice of the group elements 
$g_i$ as well as on the momenta and quantum numbers of external particles.
Moreover it factorizes on the products of certain kinematic factors (traces) and
{\it $g$-independent} amplitude $F(p_1,...p_K)$. 
The last observation is very 
crucial.
The difference between the original ${\cal N}=4$ 
and orbifold theories in 
encoded in these
kinematic factors.
Every boundary of the two-dimensional diagram 
contributes a kinematic factor -- 
trace
\eqn\trac{{\rm Tr}(\prod \gamma_{g_i})~}
for every boundary without external particles and 
\eqn\trac{{\rm Tr} \left[ \lambda_1 \left( \prod \gamma_{g_i} \right) 
\lambda_{2}
 \left( \prod \gamma_{g_j} \right) ... \lambda_{K} \right]~}
for the boundary with external particles.

Probably before going further it is instructive to consider two examples --
planar and non-planar diagrams. 
Later we formalize our arguments for the general 
case.
Consider first the planar diagram shown on Fig. 1.

\vskip .2in
\let\picnaturalsize=N
\def\picsize{2.4in}
\def\picfilename{LN1.eps}
\ifx\nopictures Y\else{\ifx\epsfloaded Y\else\fi
\global\let\epsfloaded=Y
\centerline{\ifx\picnaturalsize N\epsfxsize \picsize\fi
\epsfbox{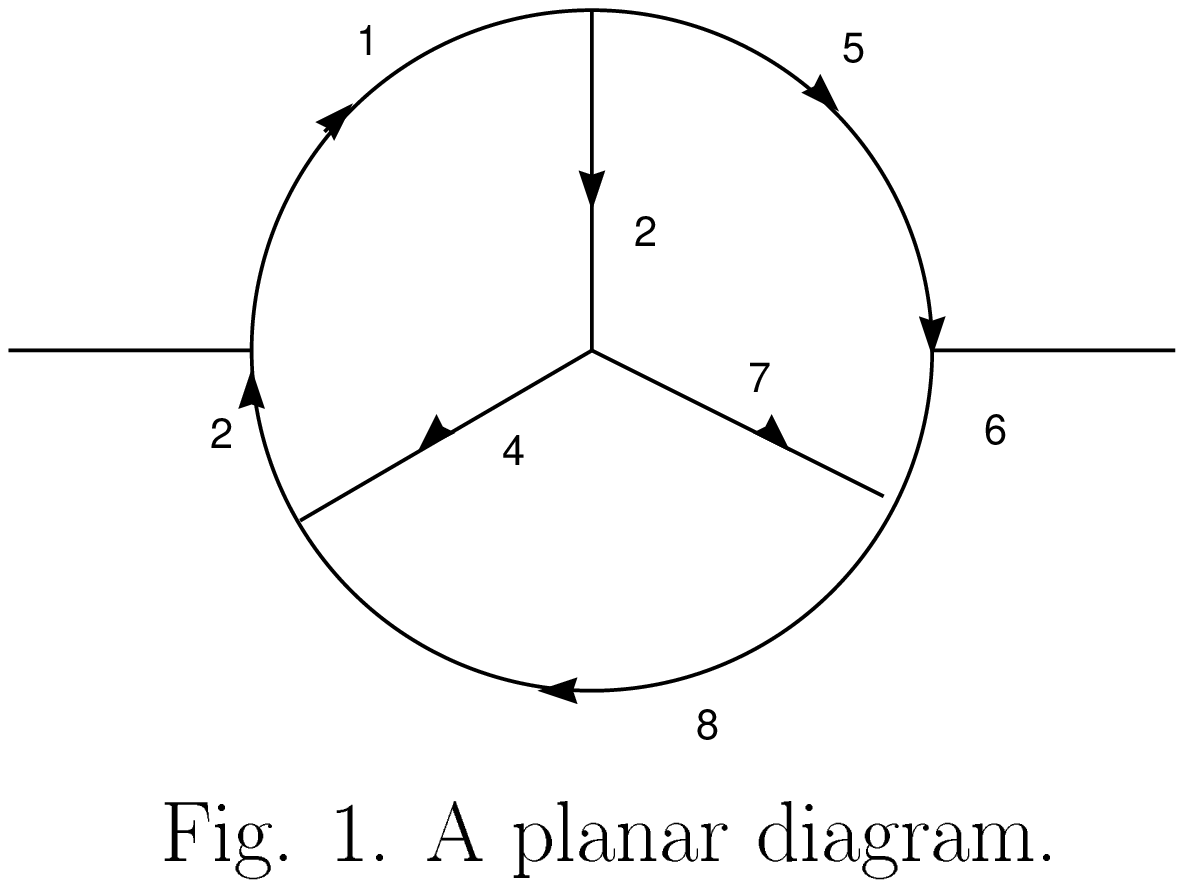}}}\fi
\vskip -.0in

In order to avoid certain 
confusions
we labeled all propagators as well as we put some arrows on them.
For diagram shown on Fig.1 we get the product of four traces
\eqn\trone{{\rm Tr}(\gamma_1 \gamma_3 \gamma_4 \gamma_2 ) ~ 
{\rm Tr}(\gamma_5 \gamma_6  \gamma_7 ^+ \gamma_3 ^+) ~
{\rm Tr}(\gamma_7 \gamma_8  \gamma_4 ^+) ~ 
{\rm Tr}(\lambda_1 \gamma_1\gamma_5  
\lambda_{2} \gamma_6  \gamma_8  \gamma_2 )}
Now let us impose the conditions that the representation is {\it regular}. That 
means that 
diagram on Fig. 1 is different from zero only in the case when 
\eqn\relone{\gamma_1 \gamma_3 \gamma_4 \gamma_2 =I~~, ~~~
\gamma_5  \gamma_6  \gamma_7 ^+ \gamma_3 ^+=I~~,~~~
\gamma_7  \gamma_8  \gamma_4 ^+=I} 
Let us take into account these relations and reexpress $\gamma_6, \gamma_8$ and 
$\gamma_2$
in terms of the remaining matrices.
After trivial manipulation we end up with 
\eqn\outb{ {\rm Tr} \left( \lambda_1 \gamma_1
\gamma_5  \lambda_2 \gamma_6  \gamma_8  \gamma_2  \right)=
{\rm Tr} \left( \lambda_1 \gamma_1 \gamma_5
\lambda_{2}  \gamma_5 ^+ \gamma_1 ^+ \right)=
{\rm Tr}(\lambda_1  \lambda_{2}  )  }
In passing to the last identity we also used the fact that all external lines 
lie in the
orbifold theory and therefore satisfy the relation 
$\gamma\lambda \gamma^+ 
=\lambda$.
Let us go back to the sum \facr. 
Instead of summing over 8 different $g_i$, we have 
to sum
only over 5 $g_i$'s and moreover the amplitude does 
not depend on the choice of 
these group elements.
Therefore we get that this amplitude is identical to ${\cal N}=4$ amplitude, 
modulo a rescaling
given by a factor $1/|\Gamma|^3$.

Let us see what happens for non-planar diagram shown on Fig.2.

\let\picnaturalsize=N
\def\picsize{2.4in}
\def\picfilename{LN2.eps}
\ifx\nopictures Y\else{\ifx\epsfloaded Y\else\fi
\global\let\epsfloaded=Y
\centerline{\ifx\picnaturalsize N\epsfxsize \picsize\fi
\epsfbox{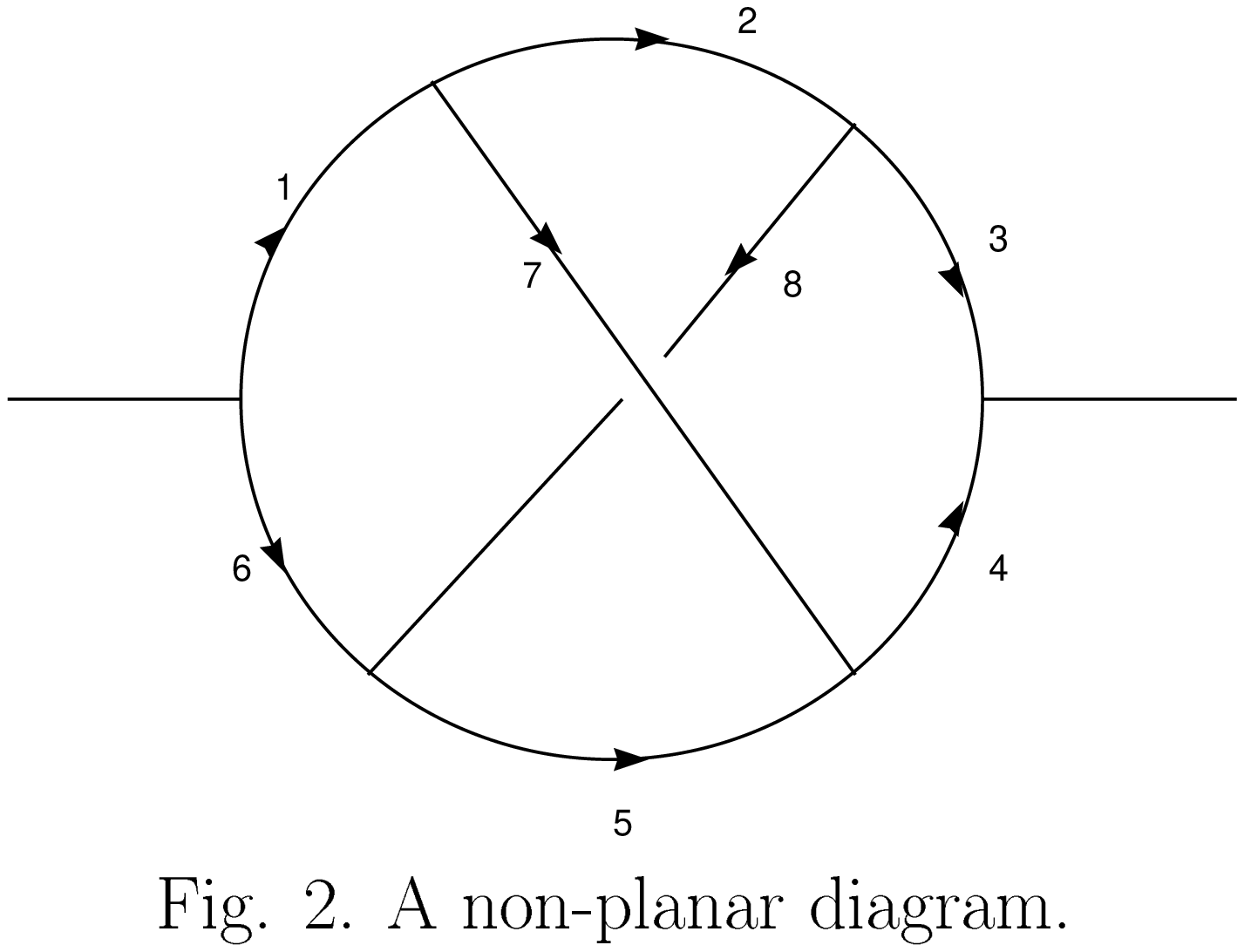}}}\fi
\vskip 0in

This diagram contains a product of two traces
\eqn\tras{{\rm Tr} \left( \gamma_1 \gamma_7 \gamma_5 ^+ \gamma_8 ^+  
\gamma_4 ^+ \gamma_7 ^+ \gamma_2 \gamma_8 \gamma_6 ^+ \right)
{\rm Tr} \left( \lambda_1 \gamma_1 \gamma_2 \gamma_3 \lambda_2 \gamma_4 ^+
\gamma_5 ^+ \gamma_6 ^+   \right) }
Again, taking into account that trace of $\gamma$ is different from zero only in 
case
when $\gamma=I$ we get the relation 
$\gamma_1 \gamma_7 \gamma_5 ^+ \gamma_8 ^+
\gamma_4 ^+ \gamma_7 ^+ \gamma_2 \gamma_8 \gamma_6 ^+ =I$. 
Substituting this 
relation into 
the other trace one gets a complicated combination of traces
\eqn\comb{{\rm Tr} \left( \lambda_1 \gamma_1 \gamma_2 \gamma_3 \lambda_2  
\gamma_4 ^+
\gamma_5 ^+ \gamma_8 ^+ \gamma_2 ^+ \gamma_7 \gamma_4 \gamma_8 \gamma_5 \gamma_7 
^+ \gamma_1 ^+   
\right)}
Even, assuming that the orbifold group is abelian this trace can not be 
rewritten as 
${\rm Tr} (\lambda_1 \lambda_2)$. 
Therefore, we end up that non-planar amplitudes in the orbifold theory are going 
to be different from ${\cal N}=4$ theory.

Now it clear what is going on. 
The product of matrices along the boundaries {\it without}
external lines can be identified with the {\it monodromy} $M=\prod \gamma_{g_i}$
along the loop. The collection of these monodromy matrices determines
a flat connection on Riemann surface (in our case disk with some number of boundaries) that represents fattened Feynman diagram. 
In case of a {\it planar} diagram the ordered product of the 
monodromy matrices is equal to one -- $\prod_k M_k = {\rm I}$. 
In case, when the group $\Gamma$ is non abelian one has to be careful in  
ordering the monodromy matrices. 
Luckily, for regular representations all the 
monodromies
along the boundaries without external legs are equal to identity and therefore the  
monodromy along the 
boundary with external legs is also trivial 
(this identical to what happen in \bkv).
Thus we obtain
\eqn\fin{\eqalign{{\rm Tr} \left[ \lambda_1  \left( \prod \gamma_{g_i} \right)  
\lambda_{2}
\left( \prod \gamma_{g_j} \right) ... \lambda_{M}  \right]= \cr
{\rm Tr} \left( \lambda_1 M_1 \lambda_{2}
M_2 ... M_{K_1}\lambda_{K} M_{k-1}^+ ...M_2 ^+ M_1 ^+ \right)=
{\rm Tr} \left( \lambda_1 \lambda_{2}...
\lambda_{K}  \right)\cr
}}

Now one can immediately see what happens. 
Instead of having a summation in \facr\ 
over 
$R$ independent group elements we will have to sum over $R-L$ independent 
group elements because there is one relation per loop (or per internal boundary).
The planar amplitudes in the orbifold theory are going to identical to those of 
${\cal N}=4$,
modulo the rescaling $1/|\Gamma|^{L}$;
we remind the reader that $L$ counts the number of loops in the planar diagram.  
Hence the parameter of perturbative expansion in the orbifold theory
is $g^2 N$ (to be compared to $g^2|\Gamma|N$ in the ${\cal N}=4$ theory).

Let us now give a more formal argument that
every {\it primitive} loop results in an extra factor $1/|\Gamma|$
in the orbifold theory as compared to the ${\cal N}=4$ theory.
By {\it primitive} loop we mean the closed loop
on the fattened Feynman diagram, where each propagator 
is replaced by a double line.
An example of such a loop is shown in Fig. 3.

\let\picnaturalsize=N
\def\picsize{2.4in}
\def\picfilename{LN3.eps}
\ifx\nopictures Y\else{\ifx\epsfloaded Y\else\fi
\global\let\epsfloaded=Y
\centerline{\ifx\picnaturalsize N\epsfxsize \picsize\fi
\epsfbox{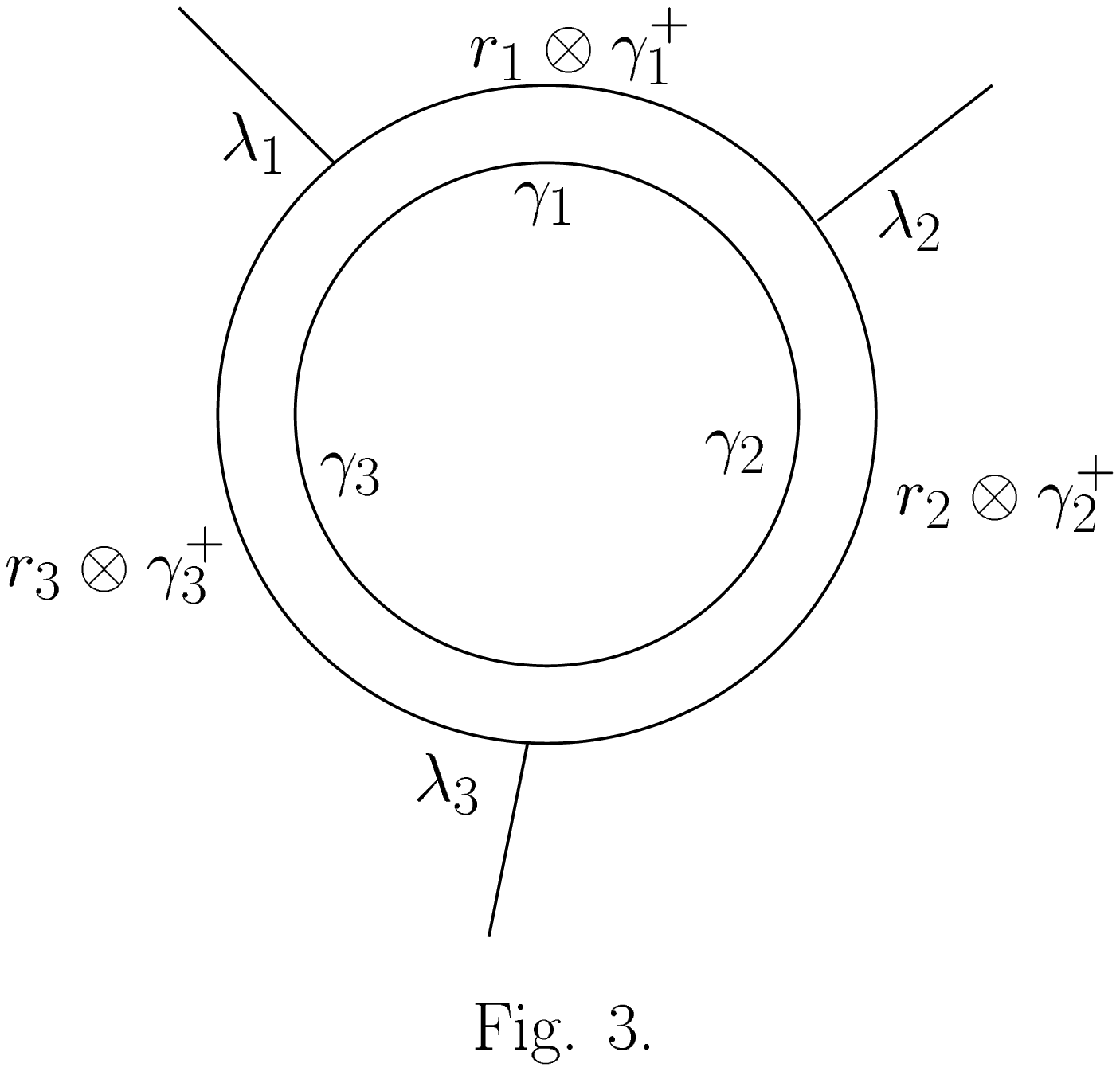}}}\fi
\vskip 0in

\noindent
Obviously, any {\it planar} Feynman diagram 
is a union of primitive loops. 
From this statement it follows that 
any planar diagram in the orbifoldized theory 
is weighted with an extra factor
$1/ |\Gamma|^{L}.$

In order to prove this statement we will use mathematical induction. The idea 
is very simple. Consider a primitive loop.
Suppose it is connected to the rest of the diagram by $m$ lines (propagators). 
Let us ``integrate'' the primitive loop out of the diagram by evaluating 
(i) momentum integral over the momentum circulating along the loop,
(ii) computing trace over indices circulating along the loop
and (iii) replacing a primitive loop by m-point  vertex. 
The $m$-point vertex operator has a very complicated momentum dependence.
This is an itterative procedure that 
decreases the number of primitive loops by one. 
The whole process stops when we integrate out all the loops
getting $M$ point correlation function.

In what follows we compare the result of integrating out a primitive loop in 
${\cal N}=4$ and ${\cal N}=2,1,0$ theories and show that they  
differ by a factor $1/|\Gamma|$ per primitive loop.
As we already mentioned the result of integrating out a loop produce a new vertex. Luckily, the momentum dependence of this new vertex is the same in both truncated and untruncated theories.
The only difference between the vertices is the dependence on the color$/$flavour indices. Therefore below we will only concentrate on the 
color$/$flavour matrices. 

There is still one small difference between original Feynman diagram and
diagrams that appear at every integration step. The original Feynman diagram 
contains only $3$- and $4$- vertices. At every integration step one 
would create new vertices with arbitrary number of external lines.
All these vertices  
are invariant with respect to R-symmetry.  
This is the only important property that we will be relevant for our discussion.

Consider a primitive loop 
in the N=4 theory with insertions of $m$ 
vertices $\{U_i\}_{i=1}^m$,
$${\rm Tr} \left( U_m\cdots U_1 \right)~,$$
where trace runs over the indices circulating in the loop.
A vertex $ U^{IJA_1...A_n}$ is a junction of 
several lines (propagators).
The indices also encodes the representations of group $\Gamma$.
We also single out two indices $I$ and $J$ corresponding  to the particle propagating in the loop. All other indices correspond to the lines connecting the primitive loop with the rest of the diagram. 
We will call these lines external
(for the primitive loop).
In order to translate this diagram into the diagram in truncated theory one has 
to make a projection of each {\it external} line
to the invariant subspace of the fields
\eqn\prov{\lambda^{IJA_1...A_n}=P_{(1)} \otimes  ... \otimes P_{(n)} ~U^{IJA_1...A_n} ~,}
where each projector $P_{(j)}$ acts on $j$-th external index $A_j$.
It follows from \invar\ that
\eqn\cym{\left(r^{(A_i)} \otimes (\gamma^+\otimes \gamma)^{(A_i)} \right) ~
\lambda^{IJA_1...A_n}=\lambda^{IJA_1...A_n}~.}

There are many ways how one can think about vertices $\lambda$. 
For fixed external indices we will think about $\lambda^{IJ...}$ as a 
{\it map} from a representation $V_I$ to $V_J$. 
This way of thinking is completely equivalent to the standard
point of view when one identifies $\lambda^{IJ...}$ with an element in
$(V_I \otimes V_J ^*)$ (for fixed external indices).
As it was mentioned above the vertex $U$ (or $\lambda$) is 
invariant under any transformation $g \in\Gamma$
simultaneously applied to all $(2+n)$ legs.
Taking \cym\ into account we get to the 
conclusion that $r_g ^{(I)} \otimes (\gamma^+\otimes \gamma)^{(I)} $
and $\lambda$ are interchangeable, namely
\eqn\prpty{
\left( r_g ^{(I)} \otimes \gamma_g ^+ \otimes \gamma_g \right)^{(I)}~
\lambda^{IJA_1...}~ = \lambda^{IJA_1...}~
\left( (r_g)^{(J)}  \otimes \gamma_g ^+  \otimes \gamma_g \right)^{(J)}~.}

Let us come back to our computation of the contribution of the 
primitive loop.
It is  clear that
the ${\cal N}=4$ (non-truncated) theory a loop fragment of a Feynman
diagram with projected {\it external} lines has
a factor of ${\rm Tr} ( \lambda_m\cdots \lambda_1)$.

In the truncated theory one has to insert the projector
$P$ into all propagators.
In this case 
a loop fragment of a planar Feynman diagram looks like
(see Fig. 3)
\eqn\loopla{\eqalign{
{\rm Tr} \left[ \prod_{i=1}^m 
                 \left( 
{1 \over | \Gamma| }  \sum_{g \in \Gamma} r_g \otimes 
\gamma^+_g \otimes \gamma_g
    \right) \lambda_i
\right] & = \cr
\left({1\over | \Gamma| }\right)^m 
\sum_{g_1,...,g_m \in \Gamma} ~
{\rm Tr} [(r_{g_1}\otimes \gamma^+_{g_1})~
\lambda_1 \cdots 
(r_{g_m}\otimes \gamma^+_{g_m})~ \lambda_m ] &
\otimes {\rm Tr} \left( \gamma_{g_1}\cdots \gamma_{g_m} \right) \cr~.
}}
Here we omitted the external indices of the vertices $\lambda$'s.
To simplify this expression we use \prpty\
a couple of times, moving all
$r_g ^{(I)} \otimes (\gamma^+\otimes \gamma)^{(I)} $, say to the right
$${\rm Tr} [(r_{g_1}\otimes\gamma^+_{g_1})~\lambda_1\cdots 
(r_{g_m}\otimes\gamma^+_{g_m})~\lambda_m]=
{\rm Tr}  \left[ \lambda_1 \cdots \lambda_m \cdot 
  (r_{g_1}\cdots r_{g_m})\otimes(\gamma_{g_1}\cdots\gamma_{g_m})
  \right].$$
Here we omitted the group elements $\otimes \gamma_{g_i}$
that are already collected in a separate 
factor ${\rm Tr} \left( \gamma_{g_1}\cdots\gamma_{g_m} \right)$
in \loopla .
The product
$\gamma_{g_1}\cdots\gamma_{g_m}$  has a nonzero trace only
if it is identically equal to $I$. This is the condition that the 
representation is regular.
It also implies that 
$g_1\cdots g_m=1$.
Therefore, the result of integrating out a primitive loop
produce am $m$ point vertex with index structure given by 
\eqn\ans{\left({1\over | \Gamma| }\right)^m
\sum_{g_1,...,g_m \in \Gamma}
{\rm Tr} \left( \lambda_1\cdots \lambda_m \right) \cdot 
{\rm Tr} \left( \gamma_{g_1}\cdots\gamma_{g_m}\right)~.}
Taking into account that there is one relation among $g_i$
and the rest of the expression is $g_i$
independent we get that sum in \ans\ gives rise to a multiplicative factor
$|\Gamma|^{m-1}$. Therefore, integrating out a primitive loop
produce a factor 
\eqn\finn{{1\over | \Gamma| } {\rm Tr} \left( \lambda_1\cdots \lambda_m\right)
 {\rm Tr} \left( {\rm I}\right)}
This is almost the ${\cal N}=4$ answer for integration out of a primitive loop,
except for the factor $1/|\Gamma|$.

Clearly, the above analysis
is also applicable to all kinds of diagrams with 
with different particles propagating in the loops. 
As the result we obtain that the scattering amplitudes in the
truncated theory 
coincides with those of ${\cal N}=4$ modulo the gauge coupling rescaling.

\vskip .2in

{\bf Acknowledgements:}

The authors thank Tom Banks, 
Zurab Kakushadze, David Kutasov, Nikita Nekrasov, Steve Shenker and Cumrun Vafa
for helpful discussions.
One of the authors also wants to thank Tom Banks, David Kutasov and Steve Shenker for encouraging to publish this work.
This work was supported by NSF grant PHY-92-18167.
The research of M.B is supported in addition by
NSF 1994 NYI award and the DOE 1994 OJI award.
M.B. would like to thank an ITP at Santa-Barbara for its hospitality,
where this work was partially done.

\listrefs

\end

One can verify that $P$ is indeed the projection operator: 
\eqn\inner{\eqalign{
P^2={1\over |\Gamma|^2}
\sum_{b\in \Gamma} \sum_{a\in \Gamma}
(r_a\otimes \gamma^+_a\otimes \gamma_a)
(r_b\otimes \gamma^+_b\otimes \gamma_b)=\cr
{1\over |\Gamma|^2}
\sum_{b\in \Gamma} \sum_{a\in \Gamma}
(r_a r_b \otimes \gamma^+_b\gamma^+_a\otimes
\gamma_a\gamma_b)=~{1\over |\Gamma|^2}
\sum_{b\in \Gamma} \sum_{c\in \Gamma}
(r_c\otimes \gamma^+_c\otimes \gamma_c)=\cr
{1\over |\Gamma|}\sum_{c\in \Gamma} r_c \otimes \gamma^+_c\otimes
\gamma_c=P
{}~.}}

Thus we have
$$[r^I\otimes (\gamma^I)^+\otimes \gamma^I] ~U
~[r^J\otimes (\gamma^J)^+\otimes \gamma^J]=
[r^K\otimes (\gamma^K)^+\otimes \gamma^K]^+ ~U
$$
Hence,
$$\lambda=\sum_{a\in \Gamma} [r_a^I\otimes (\gamma^I_a)^+\otimes
\gamma^I_a] ~U
~[r_a^J\otimes (\gamma^J_a)^+\otimes
\gamma^J_a]$$